\documentclass{article}
\usepackage{amsfonts,amssymb,amsmath}
\usepackage[dvips]{epsfig}
\usepackage[T1]{fontenc}
\usepackage[latin1]{inputenc}
\usepackage{graphicx}
\usepackage[english]{babel}
\usepackage{amsmath}
\usepackage{amssymb}
\usepackage{amsfonts}
\begin{document}

\title{\bf Induced Matter Brane Gravity and Einstein Static Universe }
\author{Y. Heydarzade$^{1}$\thanks{%
email: heydarzade@azaruniv.edu} and F. Darabi$^{1,2}$\thanks{%
email: f.darabi@azaruniv.edu; Corresponding author} ,
\\{\small $^1$Department of Physics, Azarbaijan Shahid Madani University, Tabriz, Iran}\\ {\small $^2$Research Institute for Astronomy and Astrophysics of Maragha (RIAAM), Maragha 55134-441, Iran}}
\date{\today}

\maketitle
\begin{abstract}
We investigate stability of the Einstein static universe against the scalar, vector and tensor perturbations in the context of induced matter brane gravity. It is shown that in the framework of this model, the Einstein static universe
has a positive spatial curvature. In contrast to  the classical general relativity, it is found
that a stable Einstein static universe against the scalar perturbations does exist provided that the variation of time dependent geometrical equation of state parameter is proportional to the minus of the variation
of the scale factor, $\delta \omega_{g}(t)=-C\delta a(t)$. We obtain neutral stability against the vector perturbations, and the stability against the tensor perturbations is guaranteed due to the positivity of the spatial curvature of the Einstein static universe in induced matter brane gravity.
\end{abstract}
\maketitle
\section{Introduction}
First attempts for finding a static solution of the field equations of general
relativity to describe a homogenous and isotropic universe was done by Einstein
which led him to the introduction of the cosmological constant \cite{Einstein}. Afterwards,, some works have focused on the stability of  the Einstein static universe
against scalar, vector and tensor perturbations \cite{Eddington, Harrison, Gibbons, Barrow2003, Barrow2009, Barrow2012}.
As a whole, the Einstein static universe in the Einstein general relativity is unstable
which means that it is almost impossible for the universe
to maintain its stability during a long time because of the
existence of varieties of perturbations such as the   quantum fluctuations.

A renewed motivation for studying the Einstein static universe comes from the
emergent universe scenario with the aim of   solving the initial singularity problem
of the standard model of cosmology \cite{Ellis}. In the framework of
emergent universe model, the universe is originated from an Einstein static state rather than
a big bang singularity.
Observation from WMAP7 \cite{WMAP} supports the positivity of space
curvature, in which it is found that a closed universe is
favored at the $68\%$  confidence level, and the universe
stays past-eternally in an Einstein static state and then
evolves to an inflationary phase.
However, this cosmological model suffers from a fine-tuning problem which can be ameliorated by modifications to the cosmological equations
of general relativity.
For this reason, similar static solutions have been studied in the context of the modified theories of gravity such as $f(R)$ gravity \cite{f(R)a, f(R)b, f(R)c}, $f(T)$ gravity \cite{f(T)}, Einstein-Cartan theory
\cite{Cartan}, nonconstant pressure models \cite{Pressure}, Horava-Lifshitz  gravity \cite{Horava}, Lyra geometry \cite{Darabi}, loop quantum cosmology \cite{Loop} and massive gravity \cite{massive}. Also,  this model has been explored
in braneworld scenarios inspired by string
theory  which describes gravity as a  higher-dimensional
theory so that it appears effectively 4-dimensional at low energy limits. In the
framework of the braneworld
scenarios, the standard gauge interactions are confined to the four-dimensional brane  embedded in a higher dimensional
bulk spacetime while the gravitational field
propagates into the extra dimensions \cite{Arkani, Randall, Dvali}. The Einstein static universe  studied in the framework of braneworld scenarios such as
\cite{Gergely, Zhang, Zhangb, Clarkson, Yaghoub}. As an instance, the author of \cite{Yaghoub}
considered the induced Einstein field equations on the brane which is supported by an
induced geometric energy-momentum and a confined
 matter field source on the brane. They obtained the stability conditions   in terms of constant geometric linear equation
of state parameter $\omega_{ext}=p_{extr}/\rho_{extr}$.
 Also, it is shown that for the case of radiation and matter dominated era, the stable Einstein static universe can be closed, open or flat in contrast to the vacuum energy dominated era in which the stable
Einstein static universe can not be open.


In this work, we first have a brief review on induced matter brane gravity
in which based on the Wesson induced matter
theory \cite{Wesson}, the higher dimensional spacetime (the bulk spacetime) is taken to have zero Ricci tensor.  Then, we study stability of the
Einstein static universe against the homogeneous
scalar perturbations in the context of the induced matter brane gravity.
Indeed, the induced field equations on the brane
 are studied with respect to the  perturbation in the cosmic scale factor $a(t)$ and the geometric equation of state parameter $\omega_{g}(t)$. The evolution of  the field equations are considered up to linear perturbations and all higher order terms are neglected.
Also, the units of $8\pi G=1$ are considered throughout this paper.

\section{General Geometrical Setup of the Model}
 Consider the $4D$  background manifold $\mathcal{M}_{4}$
isometrically embedded
in a $n$ dimensional bulk $\mathcal{M}_{n}$ by a differential map ${\cal Y}^{A}:\mathcal{M}_{4}\longrightarrow\mathcal{M}_{n}$ such that

\begin{eqnarray}\label{21}
{\cal G} _{AB} {\cal Y}^{A}_{,\mu } {\cal Y}^{B}_{,\nu}=
\bar{g}_{\mu \nu}  , \hspace{.5 cm} {\cal G}_{AB}{\cal
Y}^{A}_{,\mu}\bar{\cal N}^{B}_{a} = 0  ,\hspace{.5 cm}  {\cal
G}_{AB}\bar{\cal N}^{A}_{a}\bar{\cal N}^{B}_{b} = {g}_{ab},
\end{eqnarray}
where ${\cal G}_{AB}$ $(\bar{g}_{\mu \nu })$ is the metric of the bulk
(brane) space $\mathcal{M}_{n}(\mathcal{M}_{4})$ in which $\{{\cal Y}^{A}\}$ $(\{x^{\mu }\})$ is the basis
of the bulk (brane), ${\bar{\cal N}^{A}}_{a}$ are $(n-4)$ normal
unit vectors orthogonal to the brane and  $g_{ab}=\epsilon\delta_{ab}$
in which $\epsilon=\pm1$ correspond to
the two possible signature of each extra dimension. Perturbation of the background
manifold $\mathcal{M}_{4}$ in a sufficiently small
neighborhood of the brane along an arbitrary transverse direction $\xi^a$ is
given by
\begin{eqnarray}\label{22}
{\cal Z}^{A}(x^{\mu},\xi^{a}) = {\cal Y}^{A} + ({\cal
L}_{\xi}{\cal Y})^{A}, \label{eq2}
\end{eqnarray}
where ${\cal L}_{\xi^a}$ represents the Lie derivative along $\xi^a$ where
$\xi^a$
 with $a=5,...,n$ are small
parameters along ${{\cal N}^{A}}_{a}$ parameterizing the  non-compact
extra dimensions.
The presence of the tangent component of the vector $\xi$
along the brane can cause some difficulties because it can induce some undesirable
coordinate gauges. But, it is shown that in the theory of geometric perturbations, it is possible to choose this vector to be orthogonal to the background \cite{Nash}.
Then, by choosing the extra dimensions $\xi^a$ to be orthogonal to the brane, we ensure gauge
independency \cite{Jalal} and have perturbations of the embedding along the
orthogonal extra directions ${{{\cal N}}}^{A}_{a}$, giving the local
coordinates of the perturbed brane as
 \begin{eqnarray}\label{23}
&&{\cal Z}_{,\mu }^{A}(x^{\nu },\xi^a)={\cal Y}_{,\mu }^{A}(x^{\nu })+
\xi^{a}{{{\cal N}}^{A}}_{a,\mu },\nonumber\\
&& {\cal Z}_{,a }^{A}(x^{\nu },\xi^a)={{\cal N}^{A}}_{a}.
\end{eqnarray}
It is seen from equation (\ref{22}) that since the vectors ${{\cal N}}^{A}$ depend only on the local coordinates $x^{\mu }$,  ${{\cal N}}^{A}={{\cal N}}^{A}(x^\mu)$,
 so they do not
propagate along the extra dimensions \begin{equation}\label{24}
{\cal N}^{A}_{a}={\bar{\cal N}^{A}}_{\,\,\,\,\,a}+\xi^{b}\left[{\bar{\cal N}^{A}}_{a},{\bar{\cal N}^{A}}_{b}\right]={\bar{\cal N}^{A}}_{\,\,\,\,\,a}.
\end{equation}
The above  assumptions lead to the embedding equations of the
perturbed geometry
as\begin{eqnarray}\label{25}
{\cal G}_{AB}{\cal Z}_{\,\,\ ,\mu }^{A}{\cal
Z}_{\,\,\ ,\nu }^{B}=g_{\mu \nu },\hspace{0.5cm}{\cal G}_{AB}{\cal
Z}_{\,\,\ ,\mu }^{A}{\cal N}_{\,\,\ a}^{B}=g_{\mu a},\hspace{0.5cm}{\cal
G}_{AB}{\cal N}_{\,\,\
a}^{A}%
{\cal N}_{\,\,\ b}^{B}={g}_{ab}.
\end{eqnarray}
where by setting ${{{\cal N}}^{A}}_{a}={\delta^{A}}_{a}$, the metric of the bulk space ${\cal G}_{AB}$ in the Gaussian frame and in the vicinity of $\mathcal{M}_{4}$ can be
written in the following matrix form
\begin{eqnarray}\label{26}
{\cal G}_{AB}=\left( \!\!\!%
\begin{array}{cc}
g_{\mu \nu }+A_{\mu c}A_{\,\,\nu }^{c} & A_{\mu a} \\
A_{\nu b} & g_{ab}%
\end{array}%
\!\!\!\right).
\end{eqnarray}%
Then, the  line element of the bulk space
 is given by
\begin{equation}\label{27}
dS^{2}={\cal G}_{AB}d{\cal Z}^{A}d{\cal Z}^{B}=g_{\mu \nu
}(x^{\alpha },\xi^a)dx^{\mu }dx^{\nu }+g_{ab}d\xi^{a}d\xi^{b},  \end{equation}
where
\begin{eqnarray}\label{28}
g_{\mu \nu }=\bar{g}_{\mu \nu }-2\xi ^{a}\bar{K}_{\mu \nu a}+\xi
^{a}\xi ^{b}%
\bar{g}^{\alpha \beta }\bar{K}_{\mu \alpha a}\bar{K}_{\nu \beta
b},
\end{eqnarray}%
represents the metric of the perturbed brane, so that
\begin{eqnarray}\label{29}
\bar{K}_{\mu \nu a}=-{\cal G}_{AB}{\cal Y}_{\,\,\,,\mu }^{A}{\cal
N}_{\,\,\ a;\nu }^{B}=-\frac{1}{2}\frac{\partial
g_{\mu \nu }}{\partial\xi^{a} },
\end{eqnarray}%
is the extrinsic curvature of the original brane (the second
fundamental form). We use the notation $A_{\mu c}=\xi ^{d}A_{\mu
cd}$, where
\begin{equation}\label{210}
A_{\mu cd}={\cal G}_{AB}{\cal N}_{\,\,\ d;\mu }^{A}{\cal N}_{\,\,\
c}^{B}=%
\bar{A}_{\mu cd},
\end{equation}%
represents the twisting vector fields
(the normal fundamental form).  Any fixed $\xi^a$ indicates a new perturbed
brane and enables us to define an extrinsic curvature for this perturbed
brane similar to the
original one by
\begin{eqnarray}\label{211}
\tilde{K}_{\mu \nu a}=-{\cal G}_{AB}{\cal Z}_{\,\,\ ,\mu
}^{A}{\cal
N}%
_{\,\,\ a;\nu }^{B}=\bar{K}_{\mu \nu a}-\xi ^{b}\left(
\bar{K}_{\mu
\gamma a}%
\bar{K}_{\,\,\ \nu b}^{\gamma }+A_{\mu ca}A_{\,\,\ b\nu
}^{c}\right).
\end{eqnarray}
Note that the definitions (\ref{26}), (\ref{28}) and (\ref{211}) require
\begin{eqnarray}
\tilde{K}_{\mu \nu a}=-\frac{1}{2}\frac{\partial {\cal G}_{\mu
\nu
}}{%
\partial \xi ^{a}}.
\end{eqnarray}%
In geometric language, the presence of gauge fields $A_{\mu a}$
tilts the embedded family of sub-manifolds with respect to the
normal vector ${\cal N} ^{A}$. According to our construction, the
original brane is orthogonal to the normal vector ${\cal N}^{A}.$
However,  equation (\ref{25})  shows that this is not true for the
deformed geometry. Thus, we change the embedding coordinates as
\begin{eqnarray}\label{212}
{\cal X}_{,\mu }^{A}={\cal Z}_{,\mu }^{A}-g^{ab}{\cal
N}_{a}^{A}A_{b\mu },
\end{eqnarray}%
where the coordinates ${\cal X}^{A}$ describe a new family of embedded
manifolds whose members are always orthogonal to ${\cal N}^{A}$.
In this coordinates the embedding equations of the perturbed brane
is similar to the original one, described by equation (\ref{21}),
so that the coordinates ${\cal Y}^{A}$ is replaced by ${\cal X}^{A}$. This new
embedding of the local coordinates are suitable for obtaining
induced Einstein field equations on the brane. The extrinsic
curvature of a perturbed brane then becomes
\begin{eqnarray}\label{213}
K_{\mu \nu a}=-{\cal G}_{AB}{\cal X}_{,\mu }^{A}{\cal N}_{a;\nu
}^{B}=\bar{K}%
_{\mu \nu a}-\xi ^{b}\bar{K}_{\mu \gamma a}\bar{K}_{\,\,\nu
b}^{\gamma
}=-\frac{1}{2}\frac{\partial g_{\mu \nu }}{\partial \xi ^{a}},
\end{eqnarray}
which is the generalized York's relation and shows the
propagation of the extrinsic curvature as a result of the metric propagation in the direction of extra dimensions. The components of
the Riemann tensor of the bulk in the embedding vielbein
$\{{\cal X}^{A}_{, \alpha}, {\cal N}^A_a \}$, lead to the
Gauss-Codazzi equations  \cite{27}
\begin{eqnarray}\label{215}
R_{\alpha \beta \gamma \delta}=2g^{ab}K_{\alpha[ \gamma
a}K_{\delta] \beta b}+{\cal R}_{ABCD}{\cal X} ^{A}_{,\alpha}{\cal
X} ^{B}_{,\beta}{\cal X} ^{C}_{,\gamma} {\cal X}^{D}_{,\delta},
\end{eqnarray}
\begin{eqnarray}\label{216}
2K_{\alpha [\gamma c; \delta]}=2g^{ab}A_{[\gamma ac}K_{ \delta]
\alpha b}+{\cal R}_{ABCD}{\cal X} ^{A}_{,\alpha} {\cal N}^{B}_{c}
{\cal X} ^{C}_{,\gamma} {\cal X}^{D}_{,\delta},
\end{eqnarray}
where ${\cal R}_{ABCD}$ and $R_{\alpha\beta\gamma\delta}$ are the
Riemann tensors of the bulk and the perturbed brane, respectively.
One can find the Ricci tensor by contracting the Gauss equation (\ref{215}) as
\begin{eqnarray}\label{217}
R_{\mu\nu}=(K_{\mu\alpha c}K_{\nu}^{\,\,\,\,\alpha c}-K_{c} K_{\mu
\nu }^{\,\,\,\ c})+{\cal R}_{AB} {\cal X}^{A}_{,\mu} {\cal
X}^{B}_{,\nu}-g^{ab}{\cal R}_{ABCD}{\cal N}^{A}_{a}{\cal
X}^{B}_{,\mu}{\cal X}^{C}_{,\nu}{\cal N}^{D}_{b}.
\end{eqnarray}
The next contraction will give the Ricci scalar
as
\begin{equation} \label{218}
R={\cal R}+(K\circ K-K_{a}K^{a})-2g^{ab}{\cal R}_{AB}{\cal N}_{a}^{A}{\cal N}_{b}^{B}+g^{ad}g^{bc}{\cal R}_{ABCD}{\cal N}_{a}^{A}{\cal N}_{b}^{B}{\cal N}_{c}^{C}{\cal N}_{d}^{D},
\end{equation}
where we have denoted $K\circ K\equiv K_{a\mu \nu }K^{a\mu \nu }$ and $%
K_{a}\equiv g^{\mu \nu }K_{a\mu \nu }$.
Using equations
(\ref{217}) and (\ref{218}) we obtain the following relation
between the Einstein tensors of the bulk and brane
\begin{equation} \label{219}
G_{AB}{\cal X}_{,\mu }^{A}{\cal X}_{,\nu }^{B}=G_{\mu \nu }-Q_{\mu \nu }-g^{ab}
{\cal R}_{AB}{\cal N}_{a}^{A}{\cal N}_{b}^{B}g_{\mu \nu }+g^{ab}{\cal R}
_{ABCD}{\cal N}_{a}^{A}{\cal X}_{\mu }^{B}{\cal X}_{\nu }^{C}{\cal N}
_{b}^{D},
\end{equation}
where $G_{AB}$ and $G_{\mu \nu }$ are the Einstein tensors of the
bulk and brane respectively, and
\begin{equation}\label{220}
Q_{\mu \nu }=g^{ab}(K_{a\mu }^{\,\,\,\,\,\,\gamma }K_{\gamma \nu
b}-K_{a}K_{\mu \nu b})-\frac{1}{2}(K\circ K-K_{a}K^{a})g_{\mu \nu }.
\end{equation}
From the definition of $%
Q_{\mu \nu }$, it is an independent conserved
geometrical quantity as $\nabla _{\mu}Q^{\mu \nu }=0$ \cite{Maia}.

Using the decomposition of the Riemann tensor of the bulk space into the
Weyl curvature tensor, the Ricci tensor and the scalar curvature as
\begin{equation}
\mathcal{R}_{ABCD}=C_{ABCD}-\frac{2}{n-2}\left(\mathcal{G}_{B[D}\mathcal{R}_{C]A}-
\mathcal{G}_{A[D}\mathcal{R}_{C]B}\right)-
\frac{2}{(n-1)(n-2)} \mathcal{R}(\mathcal{G}_{A[D}\mathcal{R}_{C]B}),
\end{equation}
we obtain the induced $4D$ Einstein equation on the brane as
\begin{eqnarray} \label{222}
G_{\mu \nu }&=&G_{AB}{\cal X}_{,\mu }^{A}{\cal X}_{,\nu }^{B}+Q_{\mu \nu }-\mathcal{E}_{\mu\nu}+\frac{n-3}{n-2}g^{ab}
{\cal R}_{AB}{\cal N}_{a}^{A}{\cal N}_{b}^{B}g_{\mu \nu}\nonumber\\
&&- \frac{n-4}{n-2}\mathcal{R}_{AB}{\cal X}_{,\mu }^{A}{\cal X}_{,\nu }^{B}+\frac{n-4}{(n-1)(n-2)}\mathcal{R}g_{\mu\nu}, \end{eqnarray}
where ${\cal E}_{\mu \nu }=g^{ab}{\cal C}_{ABCD}{\cal X}
_{,\mu }^{A}{\cal N}_{a}^{B}{\cal N}_{b}^{C}{\cal X}_{,\nu }^{D}$
is the electric part of
the Weyl tensor of the bulk space ${\cal C}_{ABCD}$. The electric part of the Weyl tensor is well known from the brane point of view. It describes a traceless matter, denoted by dark radiation or Weyl matter.
\section{Induced Matter Brane Gravity, the Einstein Static Universe and Stability
Analysis}

Now, let us to concentrate on
the induced matter brane gravity. In this theory,
the motivation for assuming the existence of large extra
dimensions was to achieve the unification of matter and geometry,
{\it i.e.}, to obtain the properties of matter as a consequence of
extra dimensions. In the framework of the IMT, Einstein field equations in the
bulk are written in the form of
\begin{equation} \label{223}
{\cal R}_{AB}=0,
\end{equation}
where ${\cal R}_{AB}$ is the Ricci tensor of the $nD$ bulk space \cite{Wesson}.
Then, using equations (\ref{222}) and (\ref{223}), the Einstein
field equations induced on the brane become
\begin{equation}\label{ba}
G_{\mu \nu }=Q_{\mu \nu }-{\cal E}_{\mu \nu }.
\end{equation}
 Thus, from a $4D$ point of view, the empty
$nD$ field equations look like Einstein equations with induced matter source as
\begin{equation}\label{225}
8\pi G_{N}T_{\mu\nu}=Q_{\mu \nu }-{\cal E}_{\mu \nu}.
\end{equation}
In what follows, we restrict our analysis to a constant curvature bulk ($\mathcal{E}_{\mu\nu}=0$)
and will
focus on the geometrical quantity, $Q_{\mu \nu}$, as the induced matter on the brane.
As was mentioned before, $Q_{\mu \nu }$
is a conserved quantity and then  can
describe the ordinary matter fields with a
geometrical origin in accordance with the spirit of the  IMT. From this point
of view, the induced matter fields on the $4D$ brane appears as the effects
of a higher dimensional geometry.

 For the purpose of  embedding of the $FRW$ brane in a five dimensional bulk space, one should consider the metric
\begin{equation}\label{bb}
 ds^{2}=-dt^{2}+a(t)^{2}(\frac{dr^2}{1-kr^2}+r^{2}d\Omega^2),
 \end{equation}
 where $a(t)$ is the cosmic scale factor ,  $k=+1, -1$ or $0$ corresponds
 to the closed, open or flat universes and $d\Omega^2 =d\theta^2
 +sin^{2}\theta d\phi^2 $ is the metric of 2-sphere.
The
components of the extrinsic curvature for the FRW metric are given by
 \begin{eqnarray}\label{bbc}
 &&K_{00}=-\frac{1}{\dot a}\frac{d}{dt}\left(\frac{b}{a}\right),\nonumber\\
 &&K_{ij}=\frac{b}{a^2}g_{ij},\ i,j=1,2,3.
 \end{eqnarray}
where dot means derivative with respect to the cosmic time $t$ and $b=b(t)$ is an arbitrary function of it \cite{Maia}.
Then, by defining the parameters $h:=\frac{\dot b}{b}$ and $H:=\frac{\dot a}{a}$, the components of the induced geometric energy-momentum tensor on
the brane,  $Q_{\mu\nu}$,
take the form of
\begin{eqnarray}\label{bbd}
&& Q_{00}=\frac{3b^2}{a^4},\nonumber\\
&&Q_{ij}=-\frac{b^2}{a^4}\left(\frac{2h}{H}-1\right)g_{ij}.
\end{eqnarray}
Also, the geometric energy-momentum tensor $Q_{\mu\nu}$ can be identified as
a perfect fluid as
\begin{equation}\label{bbe}
 Q_{\mu\nu}=(\rho_{g} + p_{g})u_{\mu}u_{\nu}+ p_{g}g_{\mu\nu},
\end{equation}
where $u_{\alpha}=\delta^{0}_{\alpha}$ and $\rho_{g}$ and  $p_{g}$ denote the ``geometric energy density"
and ``geometric pressure", respectively (the suffix ``$g$" stands for ``geometric").
Then, using the equations  (\ref{bbd}) and (\ref{bbe}) we obtain
 \begin{eqnarray}\label{bbf}
 &&\rho_{g}=\frac{3b^2}{a^4},\nonumber\\
 &&p_{g}=-\frac{b^2}{a^4}\left(\frac{2h}{H}-1\right).
 \end{eqnarray}
In addition, the geometric fluid can be implemented by the equation of state $p_{g}=\omega_{g}\rho_{g}$ where  $\omega_{g}$ is the geometric
equation of state parameter and generally is a function of time, $\omega_{g}=\omega_{g}(t)$ . Using
 equations (\ref{bbf}) and the equation of state of the geometric fluid,
we obtain the following differential equation
 \begin{equation}\label{bbg}
\frac{\dot b}{b}=\frac{1}{2}\left(1-3\omega_{g}\right)\frac{\dot a}{a},
\end{equation}
which yields the following solution for $b(t)$ in terms of the scale factor and
the geometric equation of state parameter
 \begin{equation}\label{bbh}
b(t)=b_{0}(\frac{a}{a_{0}})^{\frac{1}{2}}exp\left(-\frac{3}{2}\int\omega_{g}(t)
\frac{\dot a}{a}dt\right),
\end{equation}
where $a_0$ and $b_0$ are the scale factor and the curvature warp of the initial Einstein static universe, respectively.

Using equations (\ref{ba}) and (\ref{bbd}), the first components of the induced Einstein equation on the brane will be
\begin{equation}\label{bbi}
\left(\frac{\dot a}{a}\right)^{2}+\frac{k}{a^2}=
\frac{b_{0}^2}{a_{0}a^3}exp\left(-3\int\omega_{g}(t)
\frac{\dot a}{a}dt\right),
\end{equation}
in which for the Einstein static universe, described by the condition $\dot a=\ddot a=0, $  gives the spatial curvature   as
\begin{equation}\label{bbj}
k=\frac{b_{0}^2}{ a_{0}^2}.
\end{equation}
 This equation
denotes that for a Einstein static universe in the framework of induced matter
brane gravity, the spatial curvature of
spacetime must be positive.

Similarly, using the equations (\ref{ba}) and (\ref{bbd}), the pressure component of the induced Einstein equation on the brane
will be\begin{equation}\label{bbk}
-2\frac{\ddot a}{a}-\left(\frac{\dot a}{a}\right)^{2}-\frac{k}{a^2}=
\frac{3b_{0}^2}{a_{0}a^3}\,\omega_{g}(t)\,exp\left(-3\int\omega_{g}(t)
\frac{\dot a}{a}dt\right).
\end{equation}
This equation, by using equation (\ref{bbj}), gives us the equation of state parameter
of the Einstein static universe as
\begin{equation}\label{bbl}
\omega_{0g}=-\frac{1}{3}.
\end{equation}

\subsection{Scalar Perturbations}
In what follows, we consider  linear homogeneous scalar perturbations of
equations (\ref{bbi}) and (\ref{bbk}) around
the Einstein static universe described by the scale factor $a_0$ and
explore its stability against these perturbations. The  perturbation in the cosmic scale factor $a(t)$ and geometric equation of state parameter
$\omega_{g}(t)$ can be considered as
\begin{eqnarray}\label{bbm}
&&a(t)\rightarrow a_{0}(1+\delta a(t)),\nonumber\\
&&\omega_{g}(t)\rightarrow \omega_{0g}(1+\delta \omega_{g}(t)).
\end{eqnarray}
Substituting these equations in equation (\ref{bbk})
and linearizing the result
gives the following equation

\begin{equation}\label{bbn}
\delta \ddot a =\frac{k}{2a_{0}^2}\delta \omega_{g},
\end{equation}
where similar process on equation (\ref{bbi}) does not yield any nontrivial
result. Equation (\ref{bbn}) represents that for having a stable Einstein static universe,
 the perturbation in the geometric equation of state parameter
 has to be in the form of $\delta \omega_{g}=-C\delta a$ where $C$ is an
arbitrary positive
constant. In this case, the equation (\ref{bbn}) has the solution
\begin{equation}\label{bbo}
\delta a=C_{1}e^{iAt}+C_{2}e^{-iAt},
\end{equation}
where $C_1$ and $C_2$ are integration constants and $A=(\frac{kC}{2a_{0}^2})^{\frac{1}{2}}$
is the frequency of oscillation around the Einstein static universe.

It is seen that the results in the framework of induced matter brane gravity are
different from the result obtained in a braneworld scenario with a confined
 matter field source on the brane \cite{Yaghoub}. In that work, it is shown that for the case of radiation and matter dominated era, the stable Einstein static universe can be closed, open or flat in contrast to the vacuum energy dominated era in which the stable
Einstein static universe can not be open. But, in the framework of induced matter
brane gravity, it is needed that the spatial curvature of the universe to
be positive and the variation of the geometric equation of state parameter
must be proportional to the minus of the variation of the scale factor.

\subsection{Vector and Tensor Perturbations}

In the cosmological context, the vector perturbations of a perfect fluid having energy density $\rho$
and barotropic pressure $p=\omega\rho$ are governed by the co-moving dimensionless {\it vorticity} defined as ${\varpi}_a=a{\varpi}$.
The vorticity modes satisfy the following propagation equation \cite{tensor}
\begin{equation}\label{v}
\dot{\varpi}_{\kappa}+(1-3c_s^2)H{\varpi}_{\kappa}=0,
\end{equation}
where $c_s^2=dp/d\rho$ and $H$  are the sound speed and the Hubble parameter,
respectively.
This equation is valid in our treatment of Einstein static universe in the
framework of the induced matter brane gravity through the
modified   Friedmann equations (\ref{bbi}) and (\ref{bbk}). For the Einstein static universe with $H=0$, equation (\ref{v}) reduces to
\begin{equation}\label{}
\dot{\varpi}_{\kappa}=0,
\end{equation}
where represents that initial vector perturbations remain frozen and consequently
we have neutral stability against vector perturbations.

Tensor perturbations, namely gravitational-wave perturbations, of a perfect
fluid is described by the co-moving dimensionless transverse-traceless shear $\Sigma_{ab}=a\sigma_{ab}$, whose modes satisfy the following equation
\begin{equation}\label{438}
\ddot\Sigma_{\kappa}+3H\dot\Sigma_{\kappa}+\left[\frac{\mathcal{K}^2}{a^2}
+\frac{2k}{a^2}-\frac{8\pi}{3}(1+3\omega)\rho\right]\Sigma_{\kappa}=0,
\end{equation}
where $\mathcal{K}$ is the co-moving index ($D^2\rightarrow -\mathcal{K}^2/a^2$ in which $D^2$ is
the covariant spatial Laplacian)\cite{tensor}.
For the Einstein static universe, this equation by
using equations (\ref{bbf}), (\ref{bbh})
(\ref{bbj}) and (\ref{bbl}),  reduces to
\begin{equation}\label{439}
\ddot\Sigma_{\kappa}+\frac{1}{ a_0^2}\left[\mathcal{K}^{2}+2k\right]\Sigma_{\kappa}=0.
\end{equation}

Then, in order to have stable modes against the tensor perturbations, the following inequality
should be satisfied
\begin{equation}
\mathcal{K}^{2}+2k>0.
\end{equation}
With respect to the result obtained in equation (\ref{bbj}), representing
the positivity of the spatial curvature of the Einstein static universe in
induced matter brane gravity, this condition is clearly satisfied. Then, in the framework of the induced matter brane gravity,  the  Einstein static universe will be stable against both of the vector and tensor perturbations.
\section{Conclusion}
Stability issue of the Einstein static universe against scalar, vector and tensor perturbations in the context of induced matter brane gravity is studied.
It is shown that in this model, the Einstein static universe
has a positive spatial curvature. The stability condition against the scalar perturbations for the Einstein static
universe appears as $\delta \omega_{g}(t)=-C\delta a(t)$ representing  the variation of the time dependent geometrical equation of state parameter in terms  of the variation of the cosmic scale factor. There is neutral stability against the vector perturbations, and the stability against the tensor perturbations is guaranteed because of the positivity of the spatial curvature of the Einstein static universe in induced matter brane gravity.
\section*{Acknowledgment}

This work has been supported financially by Research Institute for Astronomy and Astrophysics of Maragha (RIAAM) under research project NO.1/3720-4.

\end{document}